# Metasurface Broadband Solar Absorber

*Abul K. Azad[1]\*, Wilton J. M. Kort-Kamp[2,3], Milan Sykora[4], Nina R. Weisse-Bernstein[5], Ting S. Luk[6], Antoinette J. Taylor[1], Diego A. R. Dalvit[2], and Hou-Tong Chen[1]*

[1] Center for Integrated Nanotechnologies, Los Alamos National Laboratory, MS K771, Los Alamos, New Mexico 87545, USA

[2] Theoretical Division, Los Alamos National Laboratory, MS B213, Los Alamos, New Mexico 87545, USA

[3] Center for Nonlinear Studies, Los Alamos National Laboratory, Los Alamos, New Mexico 87545, USA

[4] Chemistry Division, Los Alamos National Laboratory, MS J514, Los Alamos, New Mexico 87545, USA

[5] Intelligence and Space Research Division, Los Alamos National Laboratory, MS B244, Los Alamos, New Mexico 87545, USA

[6] Center for Integrated Nanotechnologies, Sandia National Laboratories, Albuquerque, New Mexico 87123, USA

\* Electronic address: aazad@lanl.gov

ABSTRACT: We demonstrate a broadband, polarization independent, omnidirectional absorber based on a metallic metasurface architecture, which accomplishes greater than 90% absorptance in the visible and near-infrared range of the solar spectrum, and exhibits low emissivity at mid- and far-infrared wavelengths. The complex unit cell of the metasurface solar absorber consists of eight pairs of gold nano-resonators that are separated from a gold ground plane by a thin silicon dioxide spacer. Our experimental measurements reveal high-performance absorption over a wide range of incidence angles for both *s*- and *p*-polarizations. We also investigate numerically the frequency-dependent field and current distributions to elucidate how the absorption occurs within the metasurface structure. Furthermore, we discuss the potential use of our metasurface absorber design in solar thermophotovoltaics by exploiting refractory plasmonic materials.

Metamaterials have allowed the demonstration of many exotic electromagnetic phenomena and inspired some interesting potential applications [1]. While bulk metamaterials pose severe fabrication challenges particularly in the optical regime [2], planar metamaterial architectures – metasurfaces – offer alternative avenues to accomplish desirable functionalities, including the manipulation of wavefront [3], polarization conversion [4], and absorption/emission engineering [5]. Metasurface perfect absorbers with thickness much smaller than the operational wavelength are attractive in many applications such as sensing [6], compressive imaging [7], and thermal management [8, 9].

Broadband absorbers covering the entire solar spectrum are also of great interest in solar energy harvesting [10]. There have been some demonstrations of material structures as high-performance solar absorbers, for instance, using dense nanorods and nanotube films [11, 12], multilayer planar photonic structures [8, 13, 14], and photonic crystals [15]. Metasurfaces consisting of complex multi-resonator unit-cells have emerged as a powerful and flexible platform to realize multiband and broadband perfect absorption, particularly in microwave [16], terahertz [17], and infrared [18] regimes. Inspired by these earlier works, here we demonstrate the design, fabrication and characterization of a broadband omnidirectional metasurface absorber exhibiting greater than 90% absorptance in the near infrared and entire visible frequency range. The relatively simple design of the absorber allows scale up to large area fabrication using conventional nano-imprint lithography.

The schematic diagram of our metasurface absorber is illustrated in Fig. 1(a), which is based on a metal-dielectric-metal architecture. It consists of an array of 50 nm thick gold nano-resonators and a 200 nm thick gold ground plane separated by a 60 nm thick silicon dioxide spacer. A super-cell containing sixteen resonators of different sizes and shapes was employed to enable broadband absorption, and they all have four-fold symmetry to provide a polarization independent response. The design was first validated through numerical simulations using commercially available full-wave electromagnetic solvers (CST Microwave Studio and COMSOL Multiphysics). The numerical simulations were carried out using periodic boundary conditions and frequency dependent tabulated dielectric properties of gold [19] and silicon dioxide [20]. The dimensions of the resonators, their spatial distribution, and the thickness of the spacer were tuned to optimize the absorption performance within the desired spectral window. Figure 1(b) shows the simulated reflectance $R$, transmittance $T$, and absorptance $A = 1 - R - T$ under normal incidence. Due to the thick gold ground plane, the transmittance through the structure is essentially zero. The simulation reveals that our structure exhibits over 90% absorptance approximately in the $400 \text{ nm} < \lambda < 900 \text{ nm}$ spectral range, and a near-zero absorption at longer wavelengths with a cutoff wavelength $\lambda_{\text{cutoff}} \sim 1100 \text{ nm}$.

The metasurface absorber was fabricated on a silicon substrate that provides the necessary mechanical support. A 200 nm thick gold film was first deposited using electron beam evaporation, followed by chemical vapor deposition of a 60 nm thick silicon dioxide film. The array of 50 nm thick gold nano-resonators was then created using electron beam lithography, metal deposition, and a lift-off process. A scanning electron microscopy (SEM) image of the sample is shown in Fig. 2(a), where the inset depicts an expanded view of the super-cell. The image shows slight deviations from the original design due to the fabrication tolerance, as revealed by, e.g., the rounded corners of the cross resonators. The active area of the fabricated metasurface absorber is 450 μm × 450 μm. The sample was characterized at wavelengths between 350 nm and 2.5 μm using a J. A. Woollam variable angle spectroscopic ellipsometer (VASE) that allows wavelength-dependent high-accuracy reflection and transmission measurements, with the optical beam focused down to ~200 μm in diameter and for angle of incidence limited to 20°–70° due to mechanical constraints of the instrument. Sample reflection spectra were normalized to reference spectra from a gold mirror, from which we derived the sample absorptance.

The measured absorptance at 20° angle of incidence is shown in Fig. 2(b), for both *s*- and *p*-polarizations, confirming the polarization-independent high absorption over almost the entire solar spectrum. In the wavelength range $450 \text{ nm} < \lambda < 920 \text{ nm}$ our metasurface absorber accomplishes absorptance higher than 90%. The broadband absorption has a sharp edge near $\lambda_{\text{cutoff}} \sim 1100 \text{ nm}$, and the solar weighted absorptance is 88% in the wavelength range $350 \text{ nm} < \lambda < 1100 \text{ nm}$. The absorptance becomes less than 10% when $\lambda > 1250 \text{ nm}$, and at wavelengths above 1500 nm, the measured absorption is negligible (< 2%), while the simulations at the same incidence angle exhibit higher values (~4%) when using tabulated dielectric properties of gold [19]. In order to understand the origin of this discrepancy, we experimentally measured the dielectric properties of our gold film using ellipsometry and employed them in additional simulations. The results for the case of *s*-polarized incident light are shown in the inset to Fig. 2(b), revealing an improved agreement between measurements of our fabricated sample and simulations of the designed structure despite the fabrication tolerance; for *p*-polarization the agreement is similar. Taking into account in the simulations the rounded corners of the cross resonators and the slight variations of the diameters of the circles does not significantly change this good agreement, which also confirms the robustness of our metasurface absorber. The broadband and high absorption performance is maintained even when the angle of incidence increases up to 60°, as shown in the measured absorptance plotted in Figs. 2(c) and (d) for *s*- and *p*-polarized light, respectively. At these angles of incidence, we also confirmed numerically that the absorption performance remains practically unchanged while changing the sample azimuthal angle.

Our metasurface achieves broadband absorption through a combination of resonators with different geometries in a super-cell, which provides a complex dispersion that enables Fabry-Pérot destructive interference in reflection at multiple frequencies, causing light trapping and high absorption [21]. For appropriately designed structures with these frequencies sufficiently close together, broadband absorption becomes possible, particularly in the optical wavelength range where metals are lossy. To better understand the broadband nature and how energy is dissipated in our metasurface absorber, we investigated the surface current and the electric field spatial distributions. At wavelengths $\lambda > 1100$ nm, the incident light does not excite any resonators and the whole structure acts as a highly reflecting mirror; when reaching the cutoff wavelength $\lambda_{cutoff} \sim 1100$ nm the resonators start to be excited. For instance, at $\lambda \sim 1000$ nm only the larger cross resonators are excited – the distributions of the current $\boldsymbol{J}$, the electric field $\boldsymbol{E}$, and the corresponding absorption $A \propto \boldsymbol{J} \cdot \boldsymbol{E}$ are mainly concentrated within them, as shown in Fig. 3(a,d). Medium-sized resonators are excited and dominate the absorption process at intermediate wavelengths, as shown in Fig. 3(b,e) for $\lambda = 705$ nm. At shorter wavelengths all resonators contribute to the overall absorptance, and the smaller ones provide the largest contribution, as shown in Fig. 3(c,f) for $\lambda = 350$ nm. Also note that at these wavelengths the light field penetrates more into the gold ground plane, and the induced currents result in a non-negligible contribution of the ground plane to the overall absorptance.

Metasurface broadband solar absorbers can also have a profound impact in solar thermophotovoltaics (STPV) [22]. A practical STPV structure needs to operate at temperatures higher than 1000 K and, therefore, the structural and material stability becomes a critical issue for the metasurface absorber with nanoscale feature sizes. Metasurface absorbers made of appropriate structures and alternative refractory materials must be adopted in order to sustain elevated temperatures [23]. For instance, the silicon dioxide used as spacer could be replaced by zirconium dioxide, which has a much higher melting temperature. On the other hand, the nanoresonators should be made of refractory conducting materials such as tungsten, titanium nitride, or graphite. In Fig. 4 we plot the simulated absorption spectra at normal incidence of the same metasurface structure as shown in Fig. 1(a) but with gold replaced by tungsten, titanium nitride, or graphite, all of which exhibit desirable broadband and high absorption. Although the cutoff is less sharp as compared to the demonstrated gold metasurface absorber, the simulated structures reveal an improvement over some previous demonstrations in the literature [24], by exhibiting lower emissivity at mid- and far-infrared wavelengths.

In conclusion, we have demonstrated a metallic metasurface absorber that enables very high absorption over the energy-rich portion of the solar spectrum and low thermal radiation at mid- and far-infrared wavelengths. This is accomplished through a deliberately designed metasurface super-cell consisting of multiple resonators that are excited over the wavelength range of interest. Our experiments are in excellent agreement with full-wave simulations. We expect that further use of refractory materials in metasurface broadband solar absorbers will open a path toward STPV applications.

We are grateful to Z. Jacob and R. Messina for discussions. We acknowledge support from the Los Alamos National Laboratory LDRD Program. This work was performed, in part, at the Center for Integrated Nanotechnologies, a U.S. Department of Energy, Office of Basic Energy Sciences Nanoscale Science Research Center operated jointly by Los Alamos and Sandia National Laboratories. Los Alamos National Laboratory, an affirmative action/equal opportunity

**References:**

[1] W. Cai and V. Shalaev, *Optical Metamaterials: Fundamentals and Applications* (Springer-Verlag, Heidelberg, 2010).

[2] C. M. Soukoulis and M. Wegener, "Past achievements and future challenges in the development of three-dimensional photonic metamaterials," *Nat. Photon.* **5**, 523-530 (2011).

[3] N. F. Yu and F. Capasso, "Flat optics with designer metasurfaces," *Nat. Mater.* **13**, 139-150 (2014).

[4] N. K. Grady, J. E. Heyes, D. Roy Chowdhury, Y. Zeng, M. T. Reiten, A. K. Azad, A. J. Taylor, D. A. R. Dalvit, and H.-T. Chen, "Terahertz metamaterials for linear polarization conversion and anomalous refraction," *Science* **340**, 1304-1307 (2013).

[5] C. M. Watts, X. L. Liu, and W. J. Padilla, "Metamaterial electromagnetic wave absorbers," *Adv. Mater.* **24**, Op98-Op120 (2012).

[6] N. Liu, M. Mesch, T. Weiss, M. Hentschel, and H. Giessen, "Infrared perfect absorber and its application as plasmonic sensor," *Nano Lett.* **10**, 2342-2348 (2010).

[7] C. M. Watts, D. Shrekenhamer, J. Montoya, G. Lipworth, J. Hunt, T. Sleasman, S. Krishna, D. R. Smith, and W. J. Padilla, "Terahertz compressive imaging with metamaterial spatial light modulators," *Nat. Photon.* **8**, 605-609 (2014).

[8] A. P. Raman, M. A. Anoma, L. Zhu, E. Rephaeli, and S. Fan, "Passive radiative cooling below ambient air temperature under direct sunlight," *Nature* **515**, 540-544 (2014).

[9] N. N. Shi, C.-C. Tsai, F. Camino, G. D. Bernard, N. Yu, and R. Wehner, "Keeping cool: Enhanced optical reflection and radiative heat dissipation in Saharan silver ants," *Science* **349**, 298-301 (2015).

[10] H. A. Atwater and A. Polman, "Plasmonics for improved photovoltaic devices," *Nat. Mater.* **9**, 205-213 (2010).

[11] J. Q. Xi, M. F. Schubert, J. K. Kim, E. F. Schubert, M. Chen, S.-Y. Lin, W. Liu, and J. A. Smart, "Optical thin-film materials with low refractive index for broadband elimination of Fresnel reflection," *Nat. Photon.* **1**, 176-179 (2007).

[12] A. Lenert, D. M. Bierman, Y. Nam, W. R. Chan, I. Celanovic, M. Soljacic, and E. N. Wang, "A nanophotonic solar thermophotovoltaic device," *Nat. Nanotechnol.* **9**, 126-130 (2014).

[13] M. Shimizu, A. Kohiyama, and H. Yugami, "High-efficiency solar-thermophotovoltaic system equipped with a monolithic planar selective absorber/emitter," *J. Photon. Energy* **5**, 053099 (2015).

[14] H. Deng, Z. Li, L. Stan, D. Rosenmann, D. Czaplewski, J. Gao, and X. Yang, "Broadband perfect absorber based on one ultrathin layer of refractory metal," *Opt. Lett.* **40**, 2592-2595 (2015).

[15] J. B. Chou, Y. X. Yeng, Y. E. Lee, A. Lenert, V. Rinnerbauer, I. Celanovic, M. Soljacic, N. X. Fang, E. N. Wang, and S. G. Kim, "Enabling ideal selective solar absorption with 2D metallic dielectric photonic crystals," *Adv. Mater.* **26**, 8041-8045 (2014).

[16] X. Shen, T. J. Cui, J. Zhao, H. F. Ma, W. X. Jiang, and H. Li, "Polarization-independent wide-angle triple-band metamaterial absorber," *Opt. Express* **19**, 9401-9407 (2011).

[17] L. Huang, D. Roy Chowdhury, S. Ramani, M. T. Reiten, S.-N. Luo, A. J. Taylor, and H.-T. Chen, "Experimental demonstration of terahertz metamaterial absorbers with a broad and flat high absorption band," *Opt. Lett.* **37**, 154-156 (2012).

[18] X. L. Liu, T. Tyler, T. Starr, A. F. Starr, N. M. Jokerst, and W. J. Padilla, "Taming the blackbody with infrared metamaterials as selective thermal emitters," *Phys. Rev. Lett.* **107**, 045901 (2011).

[19] P. B. Johnson and R. W. Christy, "Optical constants of noble metals," *Phys. Rev. B* **6**, 4370-4379 (1972).

[20] E. D. Palik, *Handbook of Optical Constants of Solids* (Academic Press, New York, 1998).

[21] H.-T. Chen, "Interference theory of metamaterial perfect absorbers," *Opt. Express* **20**, 7165-7172 (2012).

[22] T. Bauer, *Thermophotovoltaics: Basic Principles and Critical Aspects of System Design* (Springer-Verlag, Berlin, 2011).

[23] U. Guler, A. Boltasseva, and V. M. Shalaev, "Refractory plasmonics," *Science* **344**, 263-264 (2014).

[24] W. Li, U. Guler, N. Kinsey, G. V. Naik, A. Boltasseva, J. G. Guan, V. M. Shalaev, and A. V. Kildishev, "Refractory plasmonics with titanium nitride: Broadband metamaterial absorber," *Adv. Mater.* **26**, 7959-7965 (2014).

[25] N. C. Chen, W. C. Lien, C. R. Liu, Y. L. Huang, Y. R. Lin, C. Chou, S. Y. Chang, and C. W. Ho, "Excitation of surface plasma wave at TiN/air interface in the Kretschmann geometry," *J. Appl. Phys.* **109**, 043104 (2011).

[26] S. Roberts, "Optical properties of nickel and tungsten and their interpretation according to Drudes formula," *Phys. Rev.* **114**, 104-115 (1959).

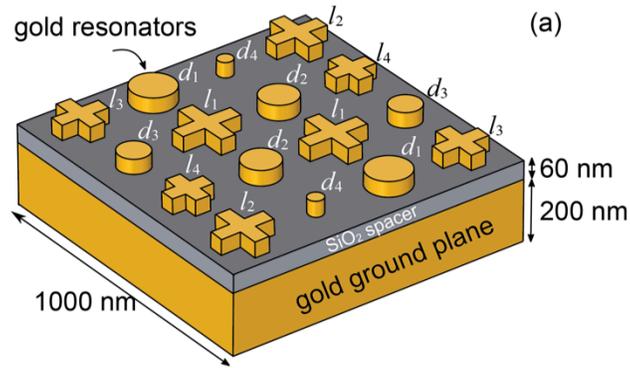

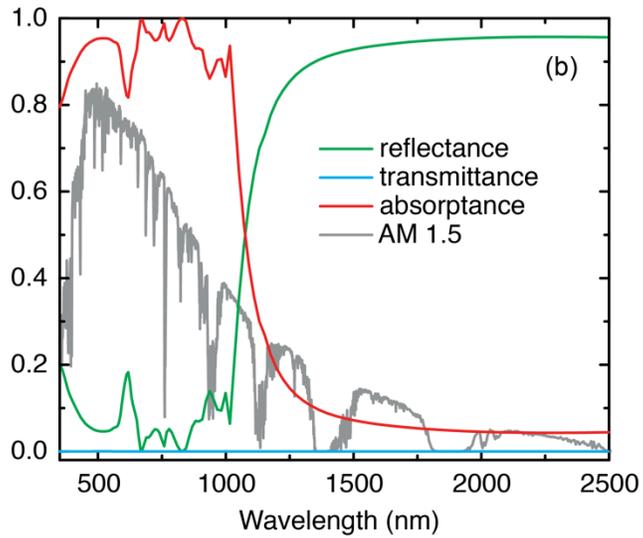

**Fig. 1:** (a) Schematic representation of the super-cell of the broadband metasurface absorber consisting of 16 resonant elements forming a square array. The lengths of the crosses are $l_1$ = 200 nm, $l_2$ = 180 nm, $l_3$ = 160 nm, and $l_4$ = 140 nm, and all crosses have identical arm widths $w$ = 50 nm. The diameters of the circles are $d_1$ = 140 nm, $d_2$ = 120 nm, $d_3$ = 100 nm, and $d_4$ = 50 nm. (b) Simulated reflectance, transmittance, and absorptance spectra at normal incidence. The gray curve shows the AM1.5 solar spectrum, normalized to fit the scale of the plot.

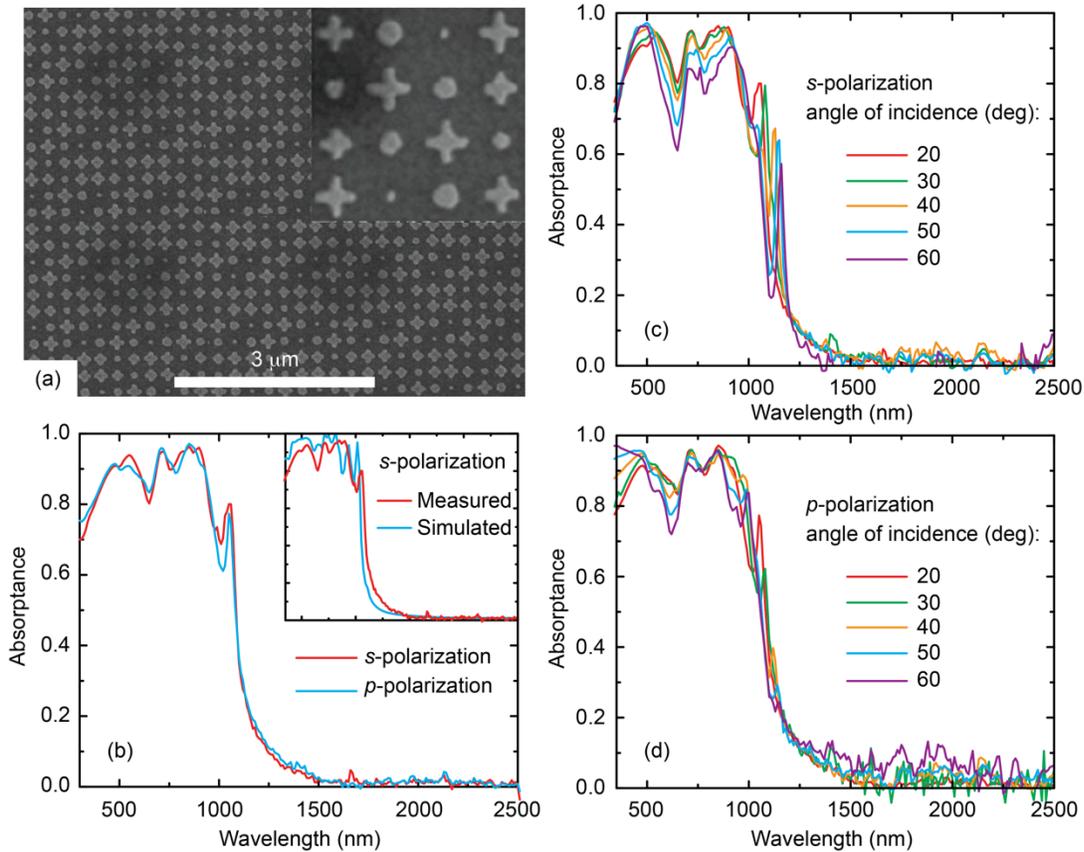

**Fig. 2:** (a) SEM image of a portion of the fabricated absorber. The inset shows an expanded view of the super-cell. (b) Experimentally measured absorptance for *s*- and *p*- polarizations at 20° angle of incidence, and at various angles of incidence for *s*- (c) and *p*-polarized (d) light. Inset to (b) is a comparison between experiments and simulations both at 20° angle of incidence for *s*-polarized incident light.

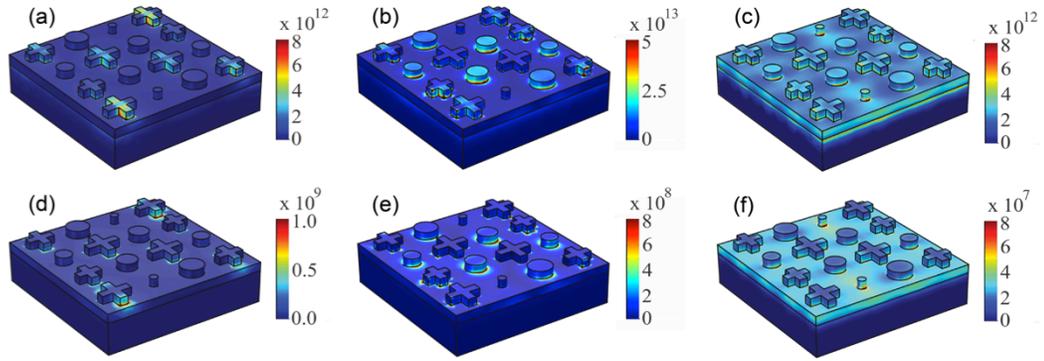

**Fig. 3:** Simulated moduli of current (top) and electric field (bottom) on the surface of the broadband metasurface absorber for impinging wavelengths of 1000 nm (a, d), 705 nm (b, e), and 350 nm (c, f). Units of current and electric field are A/m$^2$ and V/m, respectively.

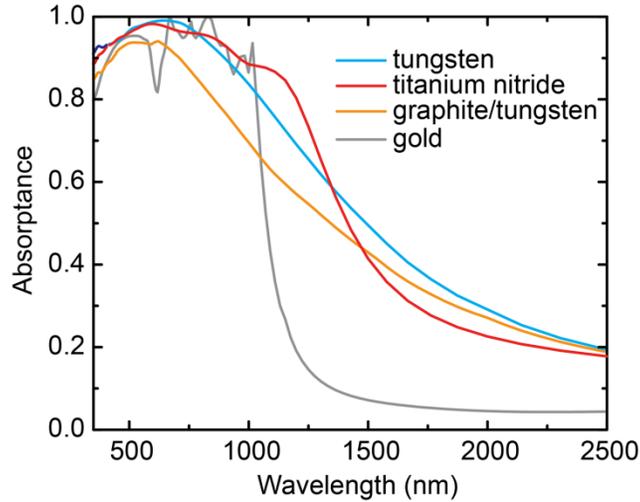

**Fig. 4:** Simulated absorption spectra at normal incidence of the metasurface absorber structure shown in Fig. 1(a), where gold is replaced by tungsten or titanium nitride in both the resonator array and the ground plane. Also shown is the case where the resonator array is made of graphite and the ground plane of tungsten. The result for the gold metasurface absorber is also shown for comparison. In all cases, silicon dioxide is used as the spacer. Numerical simulations are carried out using available optical data at room temperature for titanium nitride [25] and graphite [20], while for tungsten we use measured optical data at elevated temperatures [26].